\documentclass[conference]{IEEEtran}
\IEEEoverridecommandlockouts
\usepackage{tabularx}
\usepackage{array} % Needed to define custom column types
\newcolumntype{Y}{>{\centering\arraybackslash}X}
\usepackage{cite}
\usepackage{amsmath,amssymb,amsfonts}
\usepackage{derivative}
\usepackage{algorithmic}
\usepackage{graphicx}
\usepackage{textcomp}
\usepackage{xcolor}

\def\BibTeX{{\rm B\kern-.05em{\sc i\kern-.025em b}\kern-.08em
    T\kern-.1667em\lower.7ex\hbox{E}\kern-.125emX}}

\begin{document}

\title{Characterizing Scaling Trends of Post-Compilation Circuit Resources for NISQ-era QML Models
\thanks{Authors acknowledge funding from the EC through HORIZON-EIC-2022-PATHFINDEROPEN-01-101099697 (QUADRATURE) and HORIZON-ERC-2021-101042080 (WINC). CGA acknowledges support from the Spanish Ministry of Science, Innovation and Universities through the Beatriz Galindo program 2020 (BG20 00023) and the European ERDF under grant PID2021-123627OB-C51. EA acknowledges support from Generalitat de Catalunya, ICREA Academia Award 2024.}}

%\author{\IEEEauthorblockN{Anonymous Authors}}

\author{\IEEEauthorblockN{Rupayan Bhattacharjee, Pau Escofet, Santiago Rodrigo, Sergi Abadal, Carmen G. Almud\'ever$^{1}$, Eduard Alarc\'on}
\IEEEauthorblockA{\textit{NaNoNetworking Center in Catalonia (N3Cat), Universitat Polit\`{e}cnica de Catalunya, Spain}
\\ $^1$\textit{Universitat Polit\`ecnica de Valencia, Spain}
\\
Email: rupayan.bhattacharjee@upc.edu}
}

% \author{\IEEEauthorblockN{Rupayan Bhattacharjee}
% \IEEEauthorblockA{
% \textit{Universitat Polit\`ecnica de Catalunya}\\
% Barcelona, Spain \\
% rupayan.bhattacharjee@upc.edu}
% \and
% \IEEEauthorblockN{Pau Escofet}
% \IEEEauthorblockA{
% \textit{Universitat Polit\`ecnica de Catalunya}\\
% Barcelona, Spain \\
% pau.escofet@upc.edu}
% \and
% \IEEEauthorblockN{Santiago Rodrigo}
% \IEEEauthorblockA{
% \textit{Universitat Polit\`ecnica de Catalunya}\\
% Barcelona, Spain \\
% santiago.rodrigo@upc.edu}
% \and
% \IEEEauthorblockN{Sergi Abadal}
% \IEEEauthorblockA{
% \textit{Universitat Polit\`ecnica de Catalunya}\\
% Barcelona, Spain \\
% abadal@ac.upc.edu}
% \and
% \IEEEauthorblockN{Carmen G.\ Almud\'ever}
% \IEEEauthorblockA{
% \textit{Universitat Polit\`ecnica de Valencia}\\
% Valencia, Spain \\
% cargara2@disca.upv.es}
% \and
% \IEEEauthorblockN{Eduard Alarc\'on}
% \IEEEauthorblockA{
% \textit{Universitat Polit\`ecnica de Catalunya}\\
% Barcelona, Spain \\
% eduard.alarcon@upc.edu}
% }

\maketitle

\begin{abstract}
This work investigates the scaling characteristics of post-compilation circuit resources for Quantum Machine Learning (QML) models on connectivity-constrained NISQ processors. We analyze Quantum Kernel Methods and Quantum Neural Networks across processor topologies (linear, ring, grid, star), focusing on SWAP overhead, circuit depth, and two-qubit gate count. Our findings reveal that entangling strategy significantly impacts resource scaling, with circular and shifted circular alternating strategies showing steepest scaling. Ring topology demonstrates slowest resource scaling for most QML models, while Tree Tensor Networks lose their logarithmic depth advantage after compilation. Through fidelity analysis under realistic noise models, we establish quantitative relationships between hardware improvements and maximum reliable qubit counts, providing crucial insights for hardware-aware QML model design across the full-stack architecture.\\
\end{abstract}
\vspace{-0.2cm}

\vspace{-0.5cm}

\section{Introduction}
Quantum Machine Learning (QML) has emerged as a promising approach to leverage the unique properties of quantum systems for learning complex data distributions \cite{biamonte2017quantum}. As quantum hardware continues to improve in qubit count and fidelity, there is growing interest in developing models that can demonstrate practical quantum advantage for real-world machine learning tasks, particularly in the Noisy Intermediate-Scale Quantum (NISQ) era and beyond\cite{preskill2018quantum}.

Among the most studied paradigms are Quantum Kernel Methods (QKMs) \cite{havlivcek2019supervised} and Quantum Neural Networks (QNNs) \cite{benedetti2019parameterized, beer2020training}, where quantum circuits are used to implicitly map classical data into high-dimensional Hilbert spaces to learn complex patterns while also being suitable for NISQ devices. Although several works have demonstrated that these methods can outperform classical ones in specific regimes \cite{huang2021power, huang2022quantum, glick2024covariant, liu2021rigorous}, particularly in small-scale learning tasks \cite{cugini2023comparing, emmanoulopoulos2022quantum}, these studies provide no insight into how these methods scale, especially in terms of their implementability on near-term quantum processors with limited qubit connectivity and resources. 

Several previous approaches have attempted to address the optimal design of QML models by leveraging expressive power of Parametrized Quantum Circuits (PQCs) \cite{sim2019expressibility} and proposed robust circuit design techniques via gate quantization and pruning \cite{hu2022quantum}, error-tolerant embeddings \cite{chu2022qmlp} and quantum architecture search \cite{du2022quantum, kuo2021quantum, wang2022quantumnas}. However, most lack either their adaptability to connectivity and resource-constrained quantum processors, or an understanding of how circuit-level resources scale, particularly under realistic noise channels, impacting QML performance. 

In this work, we characterize the compilation-aware resource consumption of a wide variety of quantum circuits used in QKMs and QNNs, quantifying the resulting overhead when compiling on different processor topologies and assessing their scalability with increasing qubit counts.  Unlike prior methods, our approach allows a system architecture-aware analysis of the scalability of QML models, building groundwork for developing QML systems optimized across multiple layers of the full-stack. We further analyze the fidelity of QML models on near-term hardware while drawing insights into the optimal design of these models from a hardware-aware perspective. Our main contributions are as follows:
\begin{itemize}
    \item A systematic characterization of post-compilation resource scaling for QML models across diverse quantum processor topologies.
    \item A quantitative study relating gate error and gate time improvements to maximum reliable qubit counts for QML models.
\end{itemize}

\section{Background}

\subsection{NISQ-era QML Models}

QKMs use quantum circuits $U(\vec{x})$ acting on the state $|0\rangle^{\otimes N}$ to encode classical data $\vec{x}$ into a quantum state $|\Psi(\vec{x})\rangle$, inducing an implicit high-dimensional feature map. The resulting kernel between two data points $\vec{x_i}$ and $\vec{x_j}$ is defined as $K(\vec{x_i},\vec{x_j})= |\langle \Psi(\vec{x_i})|\Psi(\vec{x_j})\rangle|^2=|\langle0|^{\otimes N}U^\dagger(\vec{x_j})U(\vec{x_i})|0\rangle^{\otimes N}|^2$, and can be used in standard classifiers such as Support Vector Machines (SVMs). 

QNNs implement feature maps $U(\vec{x})$ to encode the data, followed by trainable PQCs (ansatz) $V(\vec{\theta})$ constructed via layered circuits of single-qubit rotations and entangling gates. The variational parameters $\vec{\theta}$ are varied to optimize the expectation value of an observable $\hat{O}$ which serves as the cost function $C(\vec{x},\vec{\theta})=\langle \Psi(\vec{x},\vec{\theta})|\hat{O}|\Psi(\vec{x},\vec{\theta})\rangle=\langle0|^{\otimes N}U^{\dagger}(\vec{x})V^{\dagger}(\vec{\theta})\hat{O}V(\vec{\theta})U(\vec{x})|0\rangle^{\otimes N}$ of the QNN. The model is trained by evaluating gradients of the cost function via the parameter-shift rule\cite{schuld2019evaluating}, i.e. $\pdv{C(\vec{x}, \vec{\theta})}{\theta_i}=\frac{1}{2}(C(\theta_i+\pi/2)-C(\theta_i-\pi/2))$.

\begin{figure}[!t]
    \centering
    \includegraphics[width=\linewidth]{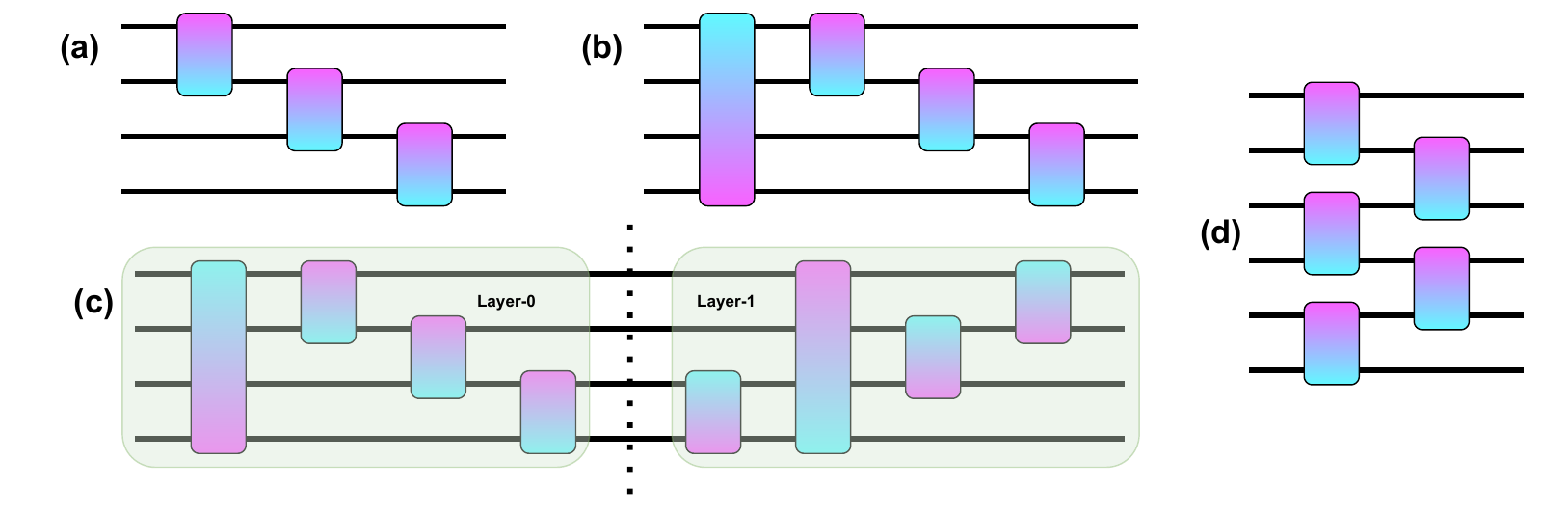}
    \vspace{-0.7cm}
    \caption{Entangling strategies with gate direction shown via color gradients: (a) Linear, each qubit connects to the next; (b) Circular, linear with an added gate between first and last qubits; (c) SCA (Shifted Circular Alternating), circular layout where, across layers, the first–last qubit connection shifts position, gate directions reverse, and application order alternates; (d) Pairwise, alternating layers of gates between odd–even and even–odd qubit pairs.}
    \label{fig:entanglement_strategies}
    \vspace{-0.5cm}
\end{figure}

The structure of the ansatz plays a key role in determining the impact of barren plateaus on the models \cite{mcclean2018barren}. The distribution of entangling two-qubit gates in the circuit, often colloquially called entangling strategy, is known to determine the expressibility and entangling capacity of QNNs \cite{sim2019expressibility}. Our work builds on this by assessing various entanglement strategies (shown in Figure \ref{fig:entanglement_strategies}) and one logarithmic-depth ansatz  under realistic hardware constraints. 

\subsection{Hardware Topologies and Compilation}

\begin{figure}[!b]
    \centering
    \vspace{-0.5cm}
    \includegraphics[width=\linewidth]{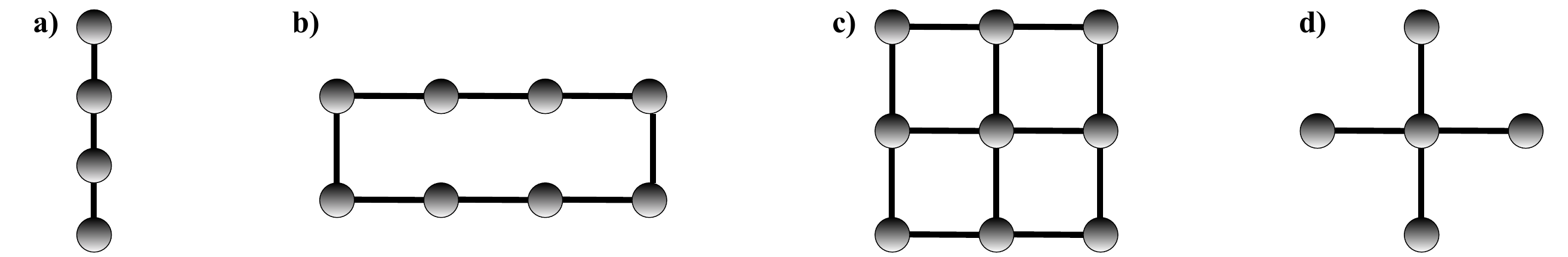}
    \vspace{-0.6cm}
    \caption{Processor topologies: (a) Linear, (b) Ring, (c) Grid, and (d) Star.}
    \label{fig:processor_topologies}
    \vspace{-0.6cm}
\end{figure}

\begin{figure*}[t]
    \centering
    \includegraphics[width=\textwidth]{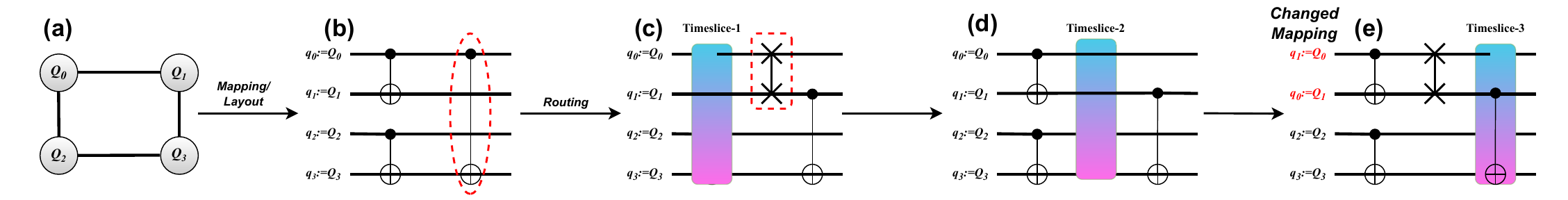}
    \vspace{-0.7cm}
    \caption{Qubit mapping and routing stages of the quantum circuit compilation process: (a) The qubit connectivity of a quantum processor (coupling map); (b) the quantum circuit to be implemented: mapping $q_i$ of the quantum circuit to $Q_i$ of the processor. It can be observed that the CNOT between $q_0$ and $q_3$ (in red) cannot be implemented as $Q_0$ and $Q_3$ are not connected; (c) routing stage: a SWAP gate is applied between $Q_0$ and $Q_1$ (in red) so that the CNOT gate can be applied between $q_0$ and $q_3$, the first layer of CNOTs is implemented in timeslice-$1$, (d) the added SWAP gate is implemented in timeslice-$2$, (e) the last CNOT gate is applied between $Q_1$ and $Q_3$ (equivalently between virtual qubits $q_0$ and $q_3$) in timeslice-3.} \vspace{-0.2cm}
    \label{fig:compilation_figure}
\end{figure*}

\begin{figure*}[t]
    \centering
    \includegraphics[width=\textwidth]{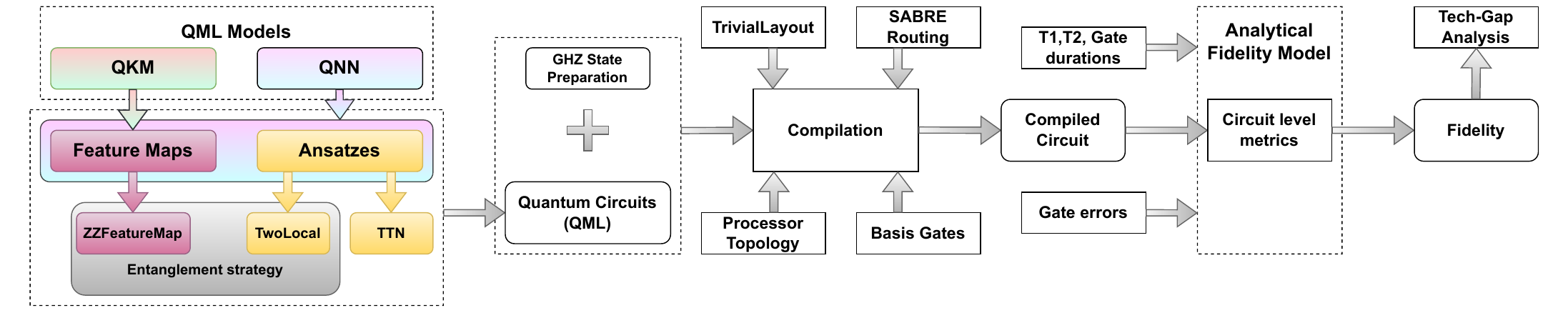}
    \vspace{-0.9cm}
    \caption{Methodology workflow.}
    \label{fig:methodology_figure}
\end{figure*}

Implementing quantum circuits on processors requires a three-stage compilation process: (1) mapping virtual qubits of the quantum circuit to physical qubits on the processor (called mapping/ layout stage), (2) inserting SWAP gates when two-qubit operations are needed between unconnected qubits (routing stage), and (3) decomposition into the processor's native gate set, and circuit optimization through gate commutations and cancellations. A visual depiction of quantum circuit compilation, particularly qubit layout and routing stages, is given in Figure \ref{fig:compilation_figure}.

The actual cost of executing any quantum circuit on a given processor topology depends on the number of two-qubit gates (i.e. qubit interactions) and their distribution in the quantum circuit. In this work, we explore the implementation cost of QML models with different two-qubit gate distributions on several hardware topologies, since the compilation overhead directly affects circuit fidelity and execution reliability.

\begin{table}[!t]
\caption{Experimental Configuration}
\centering
\begin{tabularx}{\columnwidth}{|Y|Y|}
\hline
\textbf{Parameter} & \textbf{Value} \\
\hline
\hline
Quantum Circuits & QKM (ZZFeatureMap), QNN(ZZFeatureMap+TwoLocal), TTN, GHZ. \\
\hline
Entangling strategies & `linear',`circular',`sca',`pairwise' \\
\hline
Processor topologies & Linear, Ring, Grid, Star \\
\hline
Compiler & Qiskit \cite{javadi2024quantum} \\
\hline
Mapping (Layout) & TrivialLayout \\
\hline
Routing method & SABRE \cite{li2019tackling} \\
\hline
Basis Gate Set & [$U_3(\theta,\phi,\lambda)$, CNOT] \\
\hline
Single-qubit error rate & $7.42\times 10^{-5}$ \cite{li2023error}\\
\hline
Two-qubit error rate & $7\times 10^{-4}$ \cite{negirneac2021high}\\
\hline
Single qubit gate time & $7.9$ ns \cite{hyyppa2024reducing}\\
\hline
Two-qubit gate time & $30$ ns \cite{xu2020high}\\
\hline
$T_1$ and $T_2$ & $1.2$ ms \cite{somoroff2023millisecond} and $1.16$ ms \cite{somoroff2023millisecond}\\
\hline
Qubit count ranges & TTN: [8, 16, 32,..., 1024], others: [100, 200, 300,...1000]\\
\hline
\end{tabularx}
\label{tab:experimental_configuration}
\vspace{-0.5cm}
\end{table}

\section{Methodology}

\begin{figure*}[!t]
    \centering
    \includegraphics[width=\textwidth]{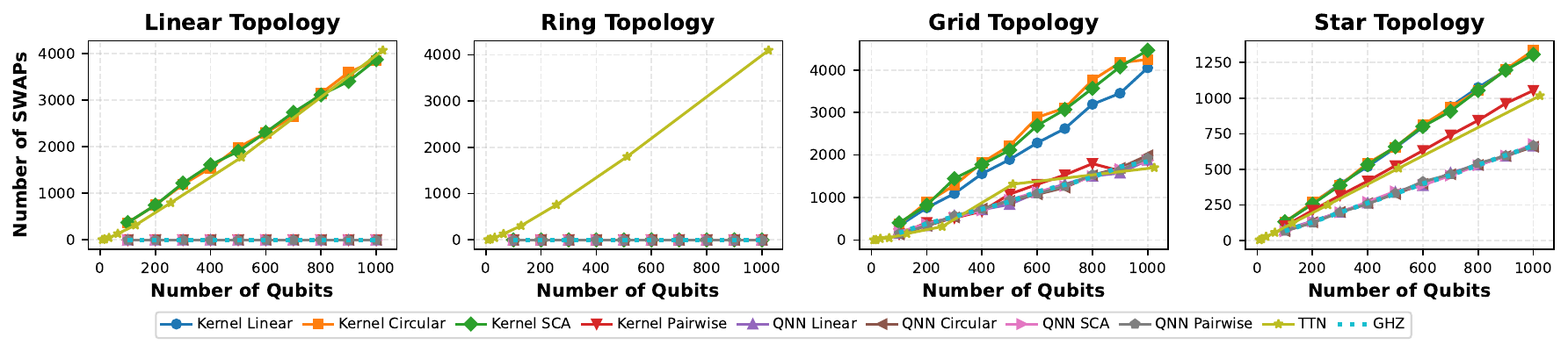}
    \vspace{-0.7cm}
    \caption{SWAP gate overhead of several QML models and GHZ state circuit upon compilation onto different processor topologies. The non-QML circuit (`GHZ') is shown as a dotted line.}
    \label{fig:SWAPs}
    \vspace{-0.5cm}
\end{figure*}

We evaluated two classes of QML models: QKMs and QNNs as they are the most NISQ-compatible. We build quantum kernel estimation circuits using the ZZFeatureMap to encode the data followed by the inverse of the same. This particular feature map is known to impart advantage in certain classification tasks \cite{havlivcek2019supervised}. We set the number of layers of ZZFeatureMap to $1$, i.e. \texttt{ZZFeatureMap(reps=1)}, and consider various circuit architectures by varying the entangling strategy. For QNN circuits, we use ZZFeatureMap with a fixed entangling strategy (`linear') followed by the TwoLocal ansatz with varying entanglement strategies. The number of layers in the feature map and the ansatz are kept at $1$ and $2$ respectively, so that the number of two-qubit gates in QKM circuits is comparable to that in the QNNs, as ZZFeatureMap has twice as many CNOTs as TwoLocal ansatz. In the following section, circuits are labeled by their type and entanglement strategy (e.g., `Kernel Circular' refers to a kernel circuit of ZZFeatureMap with circular entanglement, `QNN Pairwise' refers to ZZFeatureMap followed by TwoLocal with pairwise entanglement).

We choose the ZZFeatureMap and TwoLocal ansatz as they are the most commonly used quantum circuits in QML. Besides these, we also use the Tree Tensor Network (TTN) ansatz \cite{huggins2019towards}, as they have logarithmic circuit depth and hence unlikely to be affected by barren plateaus \cite{cerezo2021cost}. Additionally, we compare them with the GHZ state preparation circuit, as a non-QML circuit with comparable two-qubit gate complexity before compilation. For circuit resource analysis, qubit counts ranged from 100-1000 (intervals of 100) for QKM, QNN, and GHZ circuits, while TTN used 8-1024 (powers of 2) due to its binary tree structure.

\subsection{Circuit Compilation and Processor Topologies}

All quantum circuits were compiled on various quantum processor topologies using the Qiskit framework \cite{javadi2024quantum}. The circuits were decomposed to a basis gate set consisting of $U_3(\theta,\phi,\lambda)$ and $CNOT$ gates. The qubit mapping technique is TrivialLayout, which means that all the virtual qubits ($q_i$) of the circuit were mapped to the corresponding physical qubit ($Q_i$) on the processor. The qubit routing method used was SABRE \cite{li2019tackling}. The processor topologies considered were linear, ring, grid and star as shown in Figure \ref{fig:processor_topologies}. The coupling map is bidirectional, allowing the use of CNOT gates in both directions.

\subsection{Noise Model for Fidelity Estimation}

For fidelity calculations, we consider an analytical model for depolarizing noise proposed by \cite{escofet2025accurate}. This particular model has been shown to outperform all existing methods for circuit fidelity evaluations and is computationally lightweight due to its analytical nature. The compiled quantum circuit is divided into timeslices, where each timeslice represents a layer of gates that execute in parallel on the quantum processor. The fidelity of all qubits start with $1$, and, for each single-qubit gate applied, the fidelity $F_{q_i}$ of qubit $q_i$ reduces to:
\begin{equation}
    F_{q_i} \leftarrow (1-p)F_{q_i}+(1-p_{ent})\frac{p}{2}
\end{equation}
where $p$ is a depolarization parameter and $p_{ent}$ is the entanglement hyperparameter which decides the degree of correlation between the depolarizing errors in qubits. For each two-qubit gate, the fidelity of both qubits reduce to
\begin{equation}
    F_{q_i,q_j} \leftarrow \sqrt{(1-p)}F_{q_i,q_j}+(1-p_{ent})\eta,
\end{equation}
where $\eta$ is given by
\begin{equation}
    \eta=\frac{1}{2}(\sqrt{(1-p)(F_{q_i}+F_{q_j})^2+p}-\sqrt{1-p}(F_{q_i}+F_{q_j})).
\end{equation}
For every timeslice, the fidelity of each qubit is updated (including qubits with no gates applied) to capture the effects of decay and decoherence, as follows,
\begin{equation}
    F_{q_i} \leftarrow F_{q_i}.e^{-t_{layer}/T_1}.\frac{1}{2}(e^{-t_{layer}/T_2}+1),
\end{equation}
where $t_{layer}$ is the timeslice duration, set by the slowest gate time in that layer. Finally, the fidelity values of each qubit, are multiplied to yield total circuit fidelity. The single and two-qubit gate error rates, coherence times ($T_1$, $T_2$), and gate execution times are assumed to be uniform across all qubits and connections.

For technology gap analysis, we define the threshold qubit count ($N_{threshold}$) as the maximum number of qubits achievable while maintaining circuit fidelity above $0.99$. Qubit ranges of $10-100$ (interval of $10$) were used for most circuits, with TTN using $4-64$ (powers of $2$) qubits. $N_{threshold}$ values were determined by fitting stretched exponential functions to fidelity curves.

The entire experimental configuration is listed in Table \ref{tab:experimental_configuration} and the methodology workflow is depicted in Figure \ref{fig:methodology_figure}.

\begin{figure}[!t]
    \centering
    \includegraphics[width=\linewidth]{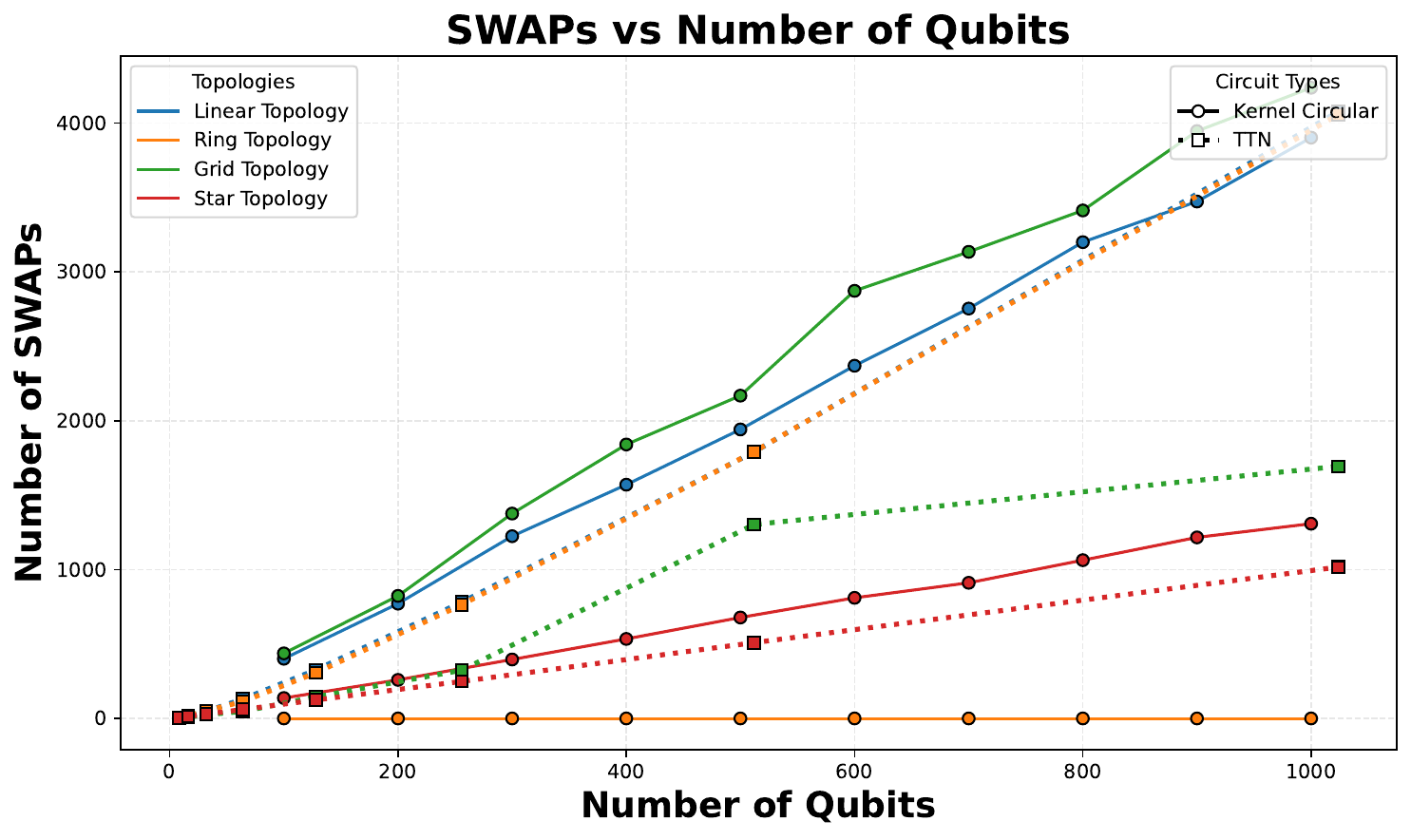}
    \vspace{-0.8cm}
    \caption{SWAP overhead scaling for Kernel Circular and TTN circuits on all processor topologies to facilitate the comparison between multiple topologies.}
    \label{fig:swap_vs_qubits}
    \vspace{-0.6cm}
\end{figure}

\section{Results and Discussion}

\begin{figure*}[!t]
    \centering
    \includegraphics[width=\textwidth]{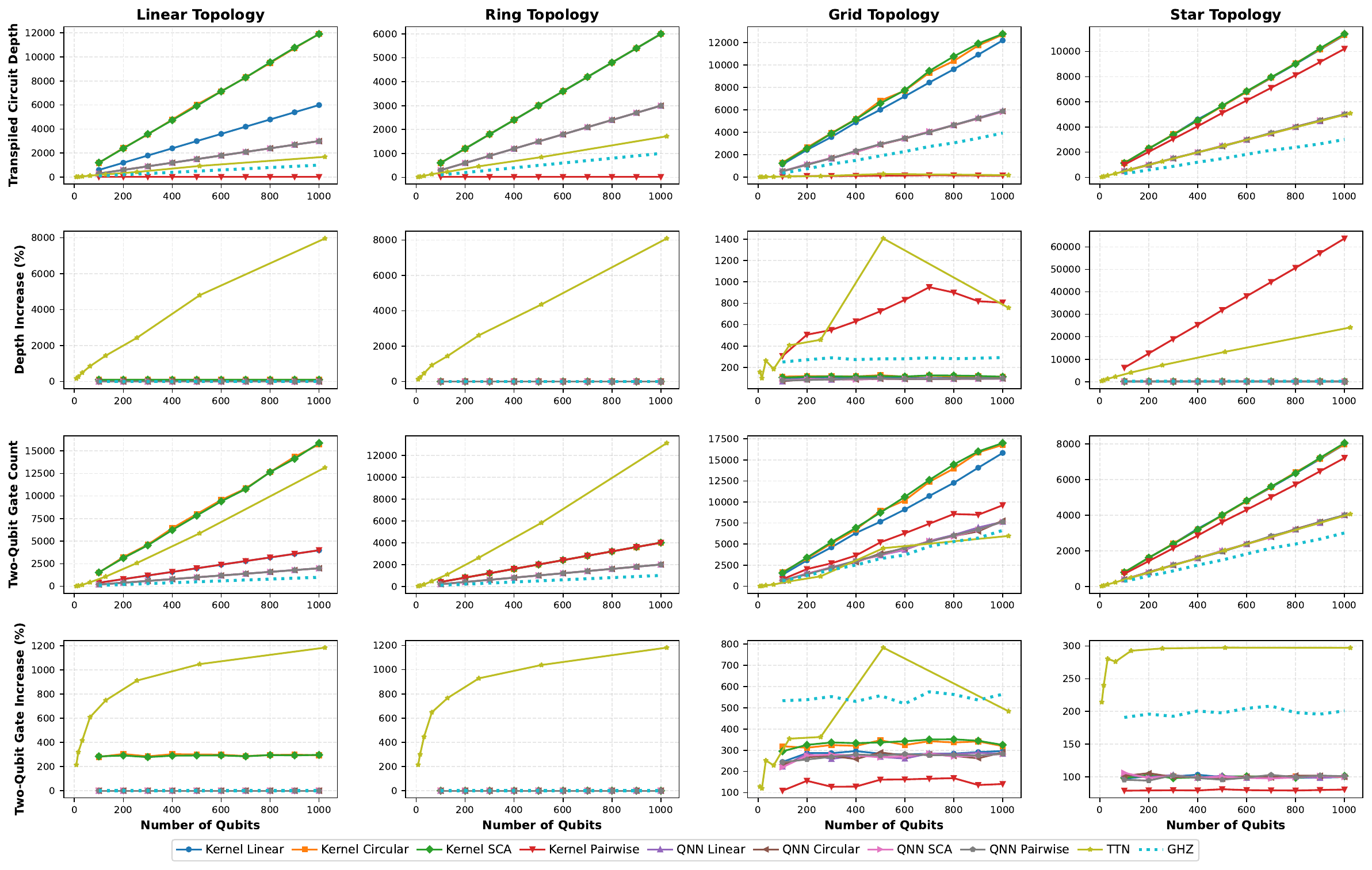}
    \vspace{-0.8cm}
    \caption{Resource scaling trends for various post-compilation circuit resources and metrics for all the circuits on all the topologies. The topologies are shown in columns and the metrics (compiled circuit depth, percentage increase in depth post compilation, two-qubit gate count after compilation and two-qubit gate overhead) are shown in rows. }
    \label{fig:all_circuit-metrics}
\end{figure*}

\begin{figure*}[!t]
    \centering
    \includegraphics[width=\textwidth]{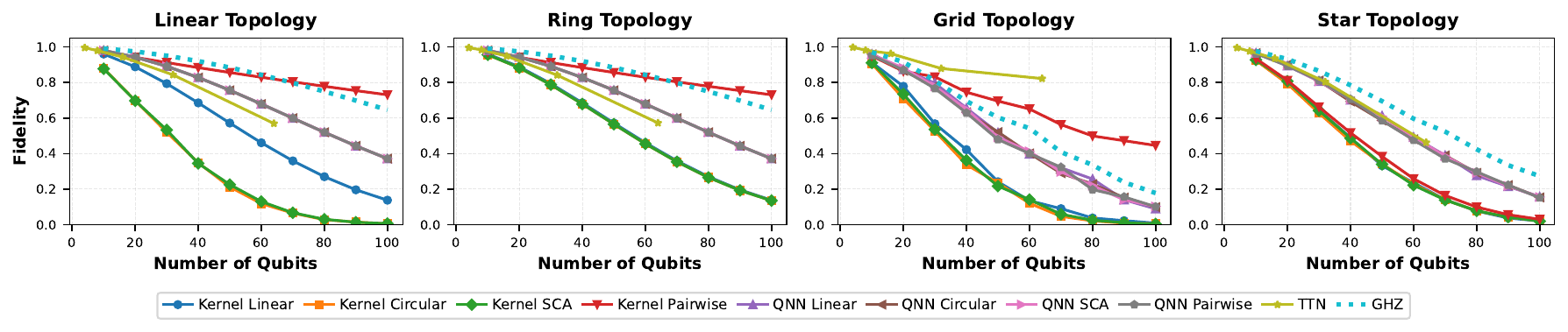}
    \vspace{-0.8cm}
    \caption{Fidelity scaling trends of QML models on various topologies with qubit counts upto a 100 qubits with state-of-the-art values for single and two-qubit gate error rates shown in Table \ref{tab:experimental_configuration}.}
    \label{fig:fidelity_vs_qubits}
    \vspace{-0.5cm}
\end{figure*}

\subsection{Circuit Resource Overhead}

Understanding post-compilation resource overhead is critical for assessing the practical scalability of QML models on near-term quantum devices with limited connectivity and for determining the reliability of circuit implementations. Figure \ref{fig:SWAPs} illustrates how the SWAP overhead scales with qubit count for various QML circuits and the GHZ state preparation circuit on different processor topologies. As shown,  all circuits follow linear trends except TTN which demonstrates a steeper scaling on linear and ring topologies. Kernel Circular  and SCA are the most expensive in terms of SWAP overhead, across all topologies except the ring topology where TTN shows the steepest scaling and all other circuits have no SWAP overhead as ring topology satisfies the qubit interactions required to implement them all. On grid and star topologies, none of the circuits can be implemented without adding SWAP gates. All QNNs and the GHZ circuit are among the least expensive to implement on all processor topologies. Overall, the optimal topology with the best SWAP overhead scaling for all models other than TTN is ring followed closely by a linear topology. For a more convenient comparison between processor topologies, we show in Figure \ref{fig:swap_vs_qubits} how the SWAP overhead scales with the number of qubits for Kernel Circular and TTN circuits. It shows that the most optimal configuration for Kernel Circular and TTN circuits are ring and star topologies respectively.

\begin{figure*}[!t]
    \centering
    \includegraphics[width=\textwidth]{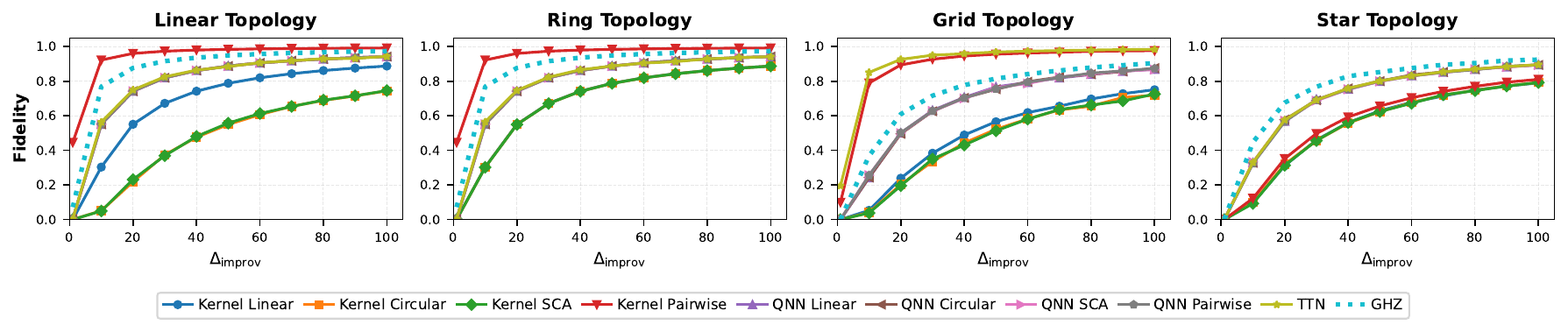}
    \vspace{-0.8cm}
    \caption{Technological Gap Analysis: fidelity vs improvement factor ($\Delta_{improv}$), for various QML models and GHZ state preparation circuit at a fixed qubit count of 256 on different processor topologies.}
    \label{fig:tech_gap_fidelity}
\end{figure*}

\begin{figure*}[!t]
    \centering
    \includegraphics[width=\textwidth]{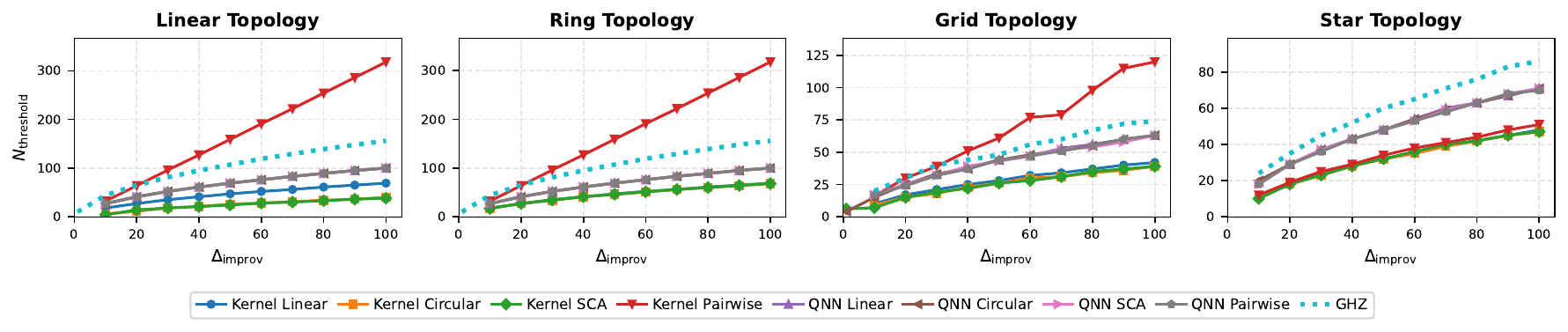}
    \vspace{-0.8cm}
    \caption{Highest qubit count with overall circuit fidelity greater than 0.99 ($N_{threshold}$) vs improvement factor ($\Delta_{improv}$), for all the circuits and topologies. TTN results omitted due to sparse and unevenly spaced sampling causing unreliable curve fitting. }
    \label{fig:tech_gap_threshold}
    \vspace{-0.5cm}
\end{figure*}

%\textcolor{red}{Note: A grid topology can always implement a quantum circuit in the same way as a linear topology does, giving the same number of circuit resource overheads. This is due to the fact that a 2D grid always consists of a linear chain. However, our results for grid and linear topologies differ as we have used a simple and computationally lightweight layout strategy i.e. TrivialLayout. More complex mapping methods can resolve this issue.}

Figure \ref{fig:all_circuit-metrics} shows various circuit-level metrics of the compiled quantum circuit on different topologies. Just as with SWAPs all circuits exhibit a linear scaling for circuit depth and two-qubit gate count, except TTN which follows a more aggressive two-qubit gate count scaling on linear and ring topologies. The Kernel Circular and SCA have the steepest scaling for compiled circuit depth and two-qubit gate overhead across all processor topologies except ring, where the worst scaling of two-qubit gates is displayed by TTN circuits. Additionally, TTN circuit that exhibits logarithmic depth scaling with the number of qubits no longer remains logarithmic but becomes linear after compilation. This behavior might affect its resilience to noise and barren plateaus \cite{cerezo2021cost}. This stresses the importance of designing processor topologies and compilation-aware circuits for greater resilience to noise and barren plateaus. The highest percentage increase in depth and two-qubit gate count is noted in TTN circuits for all but star topology. The GHZ state circuit has the lowest number of two-qubit gates in all processor topologies, making it the least affected by depolarizing gate errors. In terms of circuit depth, the Kernel Pairwise circuit is the most optimal with fixed compiled circuit depth, making it the least likely to be affected by decoherence across all topologies except star.

\subsection{Fidelity Estimation}

While circuit resource analysis provides important insights into compilation overhead, the ultimate measure of practical utility is the resulting circuit fidelity, which determines the reliability of QML computations on real quantum hardware. Figure \ref{fig:fidelity_vs_qubits} shows the fidelity of the quantum circuits for various QML models as the number of qubits increases for different quantum processor topologies. It is observed that all circuits except GHZ and Kernel Pairwise (in linear and ring topologies) encounter a significant loss of fidelity (below 0.6) at a scale of 100 qubits, rendering them highly unreliable for kernel or gradient computations. As expected, the fidelity decreases the fastest for Kernel Circular and SCA circuits. The ring topology exhibits the lowest decay in fidelity for all models except TTN which decays much slower in grid topology. In linear and ring topologies, we notice a crossover point for Kernel Pairwise and GHZ state circuits between 60 and 70 qubits after which the fidelity of Kernel Pairwise remains higher than that of the GHZ state circuit. This crossover is observed between 20 and 30 qubits in the grid topology, whereas the GHZ state circuit shows the highest fidelity in star topology where Kernel Pairwise is among the least robust circuits. All QNNs show very low fidelity (less than 0.4) at a scale of 100 qubits in all topologies. TTN circuits seem to be the most reliable in grid topology.

\subsection{Technological Gap Analysis}

To understand how technological advances impact QML model performance, we examine how improvements in gate error rates and gate times (equivalently coherence times) affect the fidelity of various QML models and the GHZ circuit. We assume that the error rates and gate times are reduced by an `Improvement Factor' ($\Delta_{improv}$) and assess the fidelity of the quantum circuits with a fixed qubit count of 256 qubits for improvement factors ranging from 1 to 100. These results are illustrated in Figure \ref{fig:tech_gap_fidelity}. For every topology, the fidelity increases rapidly with the improvement factor and then saturates. In linear and ring processor topologies, the fidelity of the Kernel Pairwise circuit saturates much faster than GHZ and other QML circuits, owing to its constant depth scaling. The exception is the grid topology, where TTN scaling is slightly faster than Kernel Pairwise. QNN circuits also converge much faster than kernel circuits, with Kernel Linear and Kernel Circular being the slowest to converge across all processor architectures, supporting the results from previous figures. All QNNs and TTN show the same scaling trend except on grid topology. The star topology exhibits the least variance between saturation curves of all circuits, but appears to be the least optimal configuration.

Additionally, Figure \ref{fig:tech_gap_threshold} shows the highest number of qubits that can be used in a given quantum circuit ($N_{threshold}$) or a QML model so that the fidelity of the overall circuit is greater than 0.99. We assess this threshold qubit count for a wide range of values of the improvement factor up to 100. It is clear that $N_{threshold}$ scales very slowly with improvement factors except for Kernel Pairwise, which demonstrates linear scaling across most processor architectures. However, even Kernel Pairwise follows slow sub-linear scaling on star topology and is among the worst performing circuits. Across all processor configurations, we notice that even with drastic technological enhancements, most models cannot reliably implement even 100 qubits, which underscores the importance of fault tolerance. However, these scaling trends would show a much higher value for $N_{threshold}$ if we set a lower target fidelity as opposed to a highly demanding value of 0.99. The non-QML GHZ circuit outperforms all QML circuits in the star topology complementing Figures \ref{fig:fidelity_vs_qubits} and \ref{fig:tech_gap_fidelity}.

\section{Conclusion}

Our comprehensive analysis of post-compilation resource scaling for QML models reveals critical insights for NISQ implementations with qubit connectivity constraints. We demonstrate that while resource requirements mostly scale linearly with qubit count across all processor topologies, ring topology exhibits the most favorable scaling characteristics for most QML models, requiring substantially fewer SWAP gates than other topologies. Importantly, entanglement strategies yield dramatically different resource requirements after compilation, with circular and SCA strategies showing the steepest scaling, while Kernel Pairwise and QNNs offer more modest requirements. A particularly significant finding is that TTN circuits lose their theoretical logarithmic depth advantage after compilation, challenging assumptions about their NISQ advantage. Our fidelity analysis shows that Kernel Pairwise circuits typically maintain higher fidelity than most circuits, with threshold qubit counts scaling linearly with gate error and gate time improvements, as opposed to other models that cannot reliably implement 100 qubits even with significant improvements. These findings allow hardware designers and algorithm developers to make informed decisions about technology requirements to scale QML implementations, while emphasizing the need for full-stack co-design approaches that consider hardware constraints during QML model design rather than relying solely on algorithm-level optimization.

\bibliographystyle{IEEEtran}
\bibliography{bibliography}

\end{document}